\documentclass[11pt,superscriptaddress,aps,prd,preprint]{revtex4}
\usepackage[active]{srcltx}
\usepackage{graphicx}
\usepackage[utf8]{inputenc}
\usepackage{amsmath}
\usepackage{verbatim}
\usepackage{xcolor}
\usepackage{subcaption}
\usepackage{times,txfonts}
\usepackage{float}

\begin{document}

\title{BTZ Black-bounce to Traversable Wormhole}

\author{J. Furtado}
\affiliation{Universidade Federal do Cariri, Juazeiro do Norte, Ceará, Brazil}
\email{job.furtado@ufca.edu.br}

\author{G. Alencar}
\affiliation{Universidade Federal do Ceará, Fortaleza, Ceará, Brazil}
\email{geova@fisica.ufc.br}

\date{\today}

\begin{abstract}

In this paper, we study the charged and uncharged BTZ counterpart of the Black-Bounce proposed by Simpson and Visser recently. For the uncharged case, we find that the temperature is not modified by the bounce parameter. We also find that the wormhole side of the solution must always be supported by exotic matter over the throat.  For the charged case we find that the thermodynamics is changed and the bounce parameter controls a phase transition, affecting the sign of the heat capacity and therefore the stability of the system. For the uncharged case, we find that there are no stable orbits for both massive and massless incoming particles while stable orbits are present for the charged case and the bounce parameter affects the points of stability.\\ 


\end{abstract}

\maketitle

\section{Introduction}\label{Sec-1}

In the last years, research in black hole physics has received renewed interest. This is due to the present era of high-precision measurements which allowed the direct measurement of gravitational waves by LIGO and VIRGO and the image of supermassive black holes by EHT \cite{LIGOScientific:2016aoc, EventHorizonTelescope:2022wkp, EventHorizonTelescope:2019dse, LIGOScientific:2017vwq}. However, the black hole interior is  riddled with a spacetime singularity. To treat this, regular black holes in $3+1$D have been proposed (see \cite{Fan:2016hvf} and references therein). A good laboratory to study black hole properties is to consider its $2+1$ counterpart, presented  in $1992$ by Bañados, Teitelboim, and Zanelli \cite{BTZ}.  The $2+1$ Einstein-Hilbert's theory at the classical level is trivial, in the sense that there are no gravitational waves, and any two solutions are locally equivalent. The $(2+1)$-dimensional vacuum has no local degrees of freedom \cite{leutwyler}. Hence, black holes were unexpected since a vacuum solution in $2+1$ dimensions is necessarily flat. However, by considering a cosmological constant with $\Lambda < 0$, Bañados, Teitelboim and Zanelli discovered the honored BTZ black holes \cite{BTZ}.

With a strictly negative cosmological constant a $2+1$, BH solution emerges presenting similar properties to the $(3+1)$-dimensional Schwarzschild and Kerr black holes. Remarkably, it admits a no-hair theorem thus characterizing the solution by its ADM-mass, angular momentum, and charge. Likewise, a rotating BTZ BH contains an inner and an outer horizon, analogous to an ergosphere in a Kerr BH. 
For our aim here, the most important characteristic of the BTZ BH is that it has thermodynamic properties closely analogous to those of realistic four-dimensional ones: a Hawking temperature and an entropy equal to a quarter of its event horizon area, where the surface area is replaced by the BTZ black hole’s disk\cite{universality,cardy}. Such notable results in a low dimensional setting strongly suggest that BH statistical mechanics in higher dimensions can be successfully investigated within $(2+1)$-dimensional gravity.

The charged BTZ black hole also presents a singularity at the origin and a regular solution has been presented in Ref. \cite{Cataldo:2000ns}. In the last years, many ways have been used to obtain regular BTZ BHs by considering non-linear sources, quasi-localized masses, and theories of modified gravity\cite{He:2017ujy,HabibMazharimousavi:2011gh,Bueno:2021krl,Guerrero:2021avm,Estrada:2020tbz,Maluf:2022ekq,Hendi:2022opt,Jusufi:2022nru}. However, very recently a new route to regular black holes has been developed and was called “black-bounce” spacetimes \cite{Simpson:2018tsi}. The name is due to the fact that the spacetime neatly interpolates between the standard Schwarzschild black hole and the Morris–Thorne traversable wormhole \cite{Morris:1988cz}. It is important to highlight that the study of traversable wormholes is a topic of great interest in the literature in several contexts \cite{Mustafa:2021vqz, Mustafa2, Tello-Ortiz:2021kxg, deSouza:2022ioq, Nilton:2022cho, Muniz:2022eex, Alencar:2021enh}. Beyond this, the geometry is regular everywhere if the “bounce” parameter $a$ is non-null. Thus, this  generalization broadness the  known  class of regular  black holes. Many properties and generalizations of the black-bounce in $(3+1)D$ have  been found, including its  rotating  and charged counterparts \cite{Mazza:2021rgq,Franzin:2021vnj,Simpson:2019cer,Lobo:2020kxn,Lobo:2020ffi,Xu:2021lff}. However, curiously, up to now only the case $D=3+1$ has been considered.

In this paper, we present the first in a series of papers in which we will study black-bounce solutions in dimensions other than $3+1$D. We will consider the BTZ black-bounce and study its properties. The manuscript is organized as follows. In section two we apply the Simpson-Visser regularization scheme to the BTZ Black Hole and analyze its properties. In section three we study the thermodynamics of the BTZ Black-bounce. In section four we analyze orbits in this spacetime. Finally, in section five, we give the conclusion.


\section{Simpson-visser BTZ Black Hole}\label{Sec-2}

The usual solution for the charged BTZ black hole is given by \cite{BTZ}
\begin{eqnarray}\label{BTZ}
ds^2=f(r)dt^2-\frac{1}{f(r)}dr^2-r^2d\phi^2;\;   f(r)=-M+\frac{r^2}{l^2}-2 q^2 \ln \left(\frac{r}{l}\right).
\end{eqnarray}
The parameter $M$ is the black hole mass, $l$ is related to the cosmological constant so that $\Lambda=-\frac{1}{l^2}$ and $q$ is the charge. The above solution has, as in the  Reissner–Nordstr\"om case, horizons and a singularity at $r=0$.  Recently, a black-bounce was proposed by Simpson and Visser. It is a way to obtain a solution that interpolates between regular black holes and a wormhole. This is controlled by a regulator parameter $a$, introduced by replacing $r\to \sqrt{r^2+a^2}$\cite{Simpson:2018tsi}. We will construct, here, a BTZ black-bounce in the same way. First, we consider the uncharged case, so that the black-bounce metric is obtained from (\ref{BTZ}) and given by
\begin{eqnarray}\label{metric}
ds^2=f(r)dt^2-\frac{1}{f(r)}dr^2-(r^2+a^2)d\phi^2; \, f(r)=\left(-M+\frac{r^2+a^2}{l^2}\right).
\end{eqnarray}

\begin{figure}[h!]
    \centering
    \includegraphics{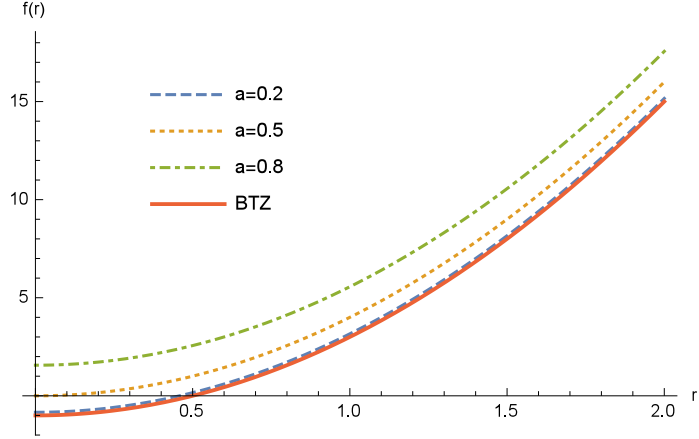}
    \caption{Modified BTZ solution. For this plot we have considered $M=1$ and $l=0.5$ in Planckian units.}
    \label{fig1}
\end{figure}

To analyze the horizons of the above solution, we note that the radial null curves are given by 
$$
\frac{dr}{dt}=\pm \left(-M+\frac{r^2+a^2}{l^2}\right)
$$
Notice that if $a^2>l^2 M$ we have $f(r)\neq0\,\,\forall\,r\,\in\,(-\infty, \infty)$ and that $dr/dt\neq 0$.  Thus we identify this region as a (2+1)-dimensions traversable wormhole\cite{Morris:1988cz}. In the case when $a^2=l^2 M$ we have $\lim_{r\rightarrow0}f(r)=0$ and that $\lim_{r\rightarrow0}dr/dt= 0$. This allows us to characterize this case as a one-way wormhole since we have an extremal null throat at $r=0$\cite{Simpson:2018tsi}. Finally, if $a^2<l^2M$ we have $dr/dt= 0$ at the location of the horizons, symmetrically placed at 
\begin{eqnarray}
r_h=\pm\sqrt{l^2 M-a^2}.
\end{eqnarray}
Therefore, the relation between the mass $M$ and the horizon radius is modified to   
\begin{eqnarray}
M=\frac{a^2+r_h^2}{l^2}.
\end{eqnarray}
The above expressions recover the usual BTZ relation when $a\rightarrow 0$. Since $a$ is a real parameter the possibility of negative mass regions in this context is disregarded. Such negative mass regions are related to black holes with non-trivial topology \cite{Mann:1997jb}. The behaviour of the Simpson-Visser BTZ solution is depicted in fig. (\ref{fig1}). 

In order to properly discuss the singularities in the present context, we calculate the curvature invariants. The Ricci scalar can be straightforwardly calculated from the metric, which gives us
\begin{equation}\label{ricci}
    R=-\frac{2 \left[2 a^4+a^2 \left(5 r^2-l^2 M\right)+3 r^4\right]}{l^2 \left(a^2+r^2\right)^2}.
\end{equation}
Note that when $a\rightarrow0$ (or alternatively $r\rightarrow\infty$) we have $R=-\frac{6}{l^2}$, which is the usual result for the BTZ black hole. Also, let us highlight that eq.(\ref{ricci}) is always finite, as expected from a regular solution. 

The squared Ricci, computed by the contraction $R_{\mu\nu}R^{\mu\nu}$ is written as
\begin{eqnarray}
    R_{\mu\nu}R^{\mu\nu}=\frac{2 \left[3 a^8+a^6 \left(14 r^2-3 l^2 M\right)+a^4 \left(l^4 M^2-7 l^2 M r^2+25 r^4\right)-4 a^2 \left(l^2 M r^4-5 r^6\right)+6 r^8\right]}{l^4 \left(a^2+r^2\right)^4},
\end{eqnarray}
while the Kretschmann scalar can be obtained by the contraction $R_{\mu\nu\lambda\rho}R^{\mu\nu\lambda\rho}$ yielding
\begin{eqnarray}
    R_{\mu\nu\lambda\rho}R^{\mu\nu\lambda\rho}=\frac{4 \left(2 a^8+a^6 \left(8 r^2-2 l^2 M\right)+a^4 \left(l^4 M^2-4 l^2 M r^2+13 r^4\right)-2 a^2 \left(l^2 M r^4-5 r^6\right)+3 r^8\right)}{l^4 \left(a^2+r^2\right)^4}.
\end{eqnarray}
Both the squared Ricci and the Kretschmann scalar are always finite independent of the value of $r$ and recover the usual BTZ results in the limit when $a=0$.

Some comments about the causal structure are necessary. First of all, for the black hole case ($a^2<Ml^2$), the Penrose diagram is identical to the one with $a=0$. The reason is that the only influence of $a$ is to change the radius of the horizon.  We also have that all the curvature tensors are the same at $r=0$.  Therefore, the causal structure is the same. The only case worth mentioning is the one-way wormhole ( $a^2=Ml^2$).  In this case, we have that $f=r^2/l^2$.  Curiously, this is exactly equal to the standard BTZ black hole with $M=0$. The Penrose diagram is given in \cite{Banados:1992gq}.  The diagram has a null throat at $r=0$. The only difference with the Schwarzschild case is that the infinity is now timelike and represented by a vertical line.

\section{Energy conditions}

We will now analyze the energy conditions for the matter supporting the Simpson-Visser black hole solution. Due to the bounce regulator parameter $a$, we must find the regions where the Null Energy Conditions (NEC, $\rho + p_i\geq0$), Weak Energy Conditions (WEC, $\rho\geq0,\, \rho + p_i\geq 0$), Strong Energy Conditions (SEC, $\rho + p_i\geq 0,\, \rho + \sum p_i\geq 0$), and Dominant Energy Conditions (DEC, $\rho \geq 0,\, -\rho \geq p_i \geq \rho$) are satisfied. The index $i$ is for the  coordinates, $(1, 2) \equiv (r, \phi)$. 

The energy density $\rho$ and the quantities $\rho + p_r$ and $\rho + p_\phi$ are given by
\begin{eqnarray}
    \rho&=&-\frac{1}{l^2}+\frac{a^2 M}{\left(a^2+r^2\right)^2},\\
    \rho+p_r&=&-\frac{a^2 \left(a^2-l^2 M+r^2\right)}{l^2 \left(a^2+r^2\right)^2}\\
    \rho+p_\phi&=&\frac{a^2 M}{\left(a^2+r^2\right)^2}.
\end{eqnarray}
Where $p_r$ and $p_\phi$ stands for the radial pressure and lateral pressure respectively. Let us notice initially that $\rho+p_\phi$ is positive everywhere for any non-vanishing bounce parameter $a$. Also, in order for $\rho+p_r\geq0$ (condition present in NEC, WEC and SEC), it is required that $r\leq l^2M-a^2$. Moreover, DEC and WEC require that $\rho\geq0$, which implies in $r<\sqrt{a l\sqrt{M}-a^2}$.  

\begin{figure*}[h!]
    \centering
\begin{subfigure}{.5\textwidth}
  \centering
  \includegraphics[scale=0.8]{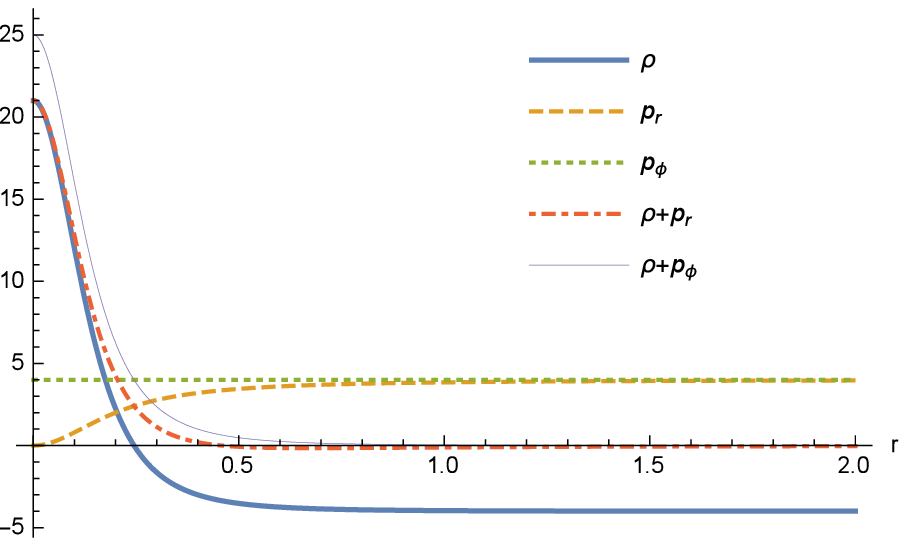}
  \caption{}

\end{subfigure}%
\begin{subfigure}{.5\textwidth}
  \centering
  \includegraphics[scale=0.8]{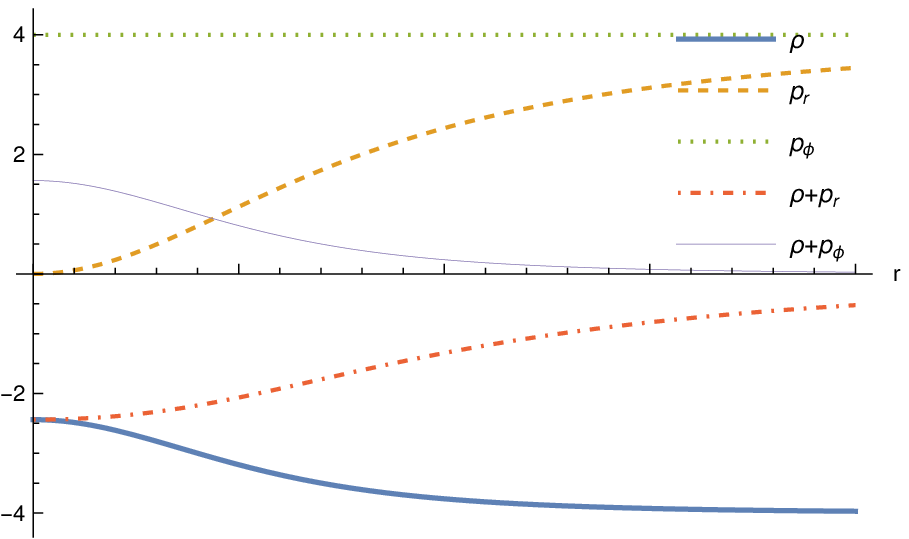}
  \caption{}

\end{subfigure}%

    \caption{Behaviour of the energy density, pressures and their combination for the Simpson-Visser BTZ black hole. The left panel stands for $a=0.2$ while the right panel is for $a=0.8$. We have considered for this plot $M=1$ and $l=0.5$.}
    \label{fig8}
\end{figure*}

For short, NEC and SEC are satisfied when $r\leq l^2M-a^2$, while WEC and DEC are satisfied when $r<\sqrt{a l\sqrt{M}-a^2}$. In fig. (\ref{fig8}) we can see the behaviour of the energy density, pressures and their combination for the present model. The left panel stands for $a=0.2$, which means that we are in the black hole domain, since we are considering $M=1$ and $l=0.5$. For the right panel we have considered the same parameters but with $a=0.8$, so that in this case we are in the wormhole domain. 

\begin{figure}[h!]
    \centering
    \includegraphics{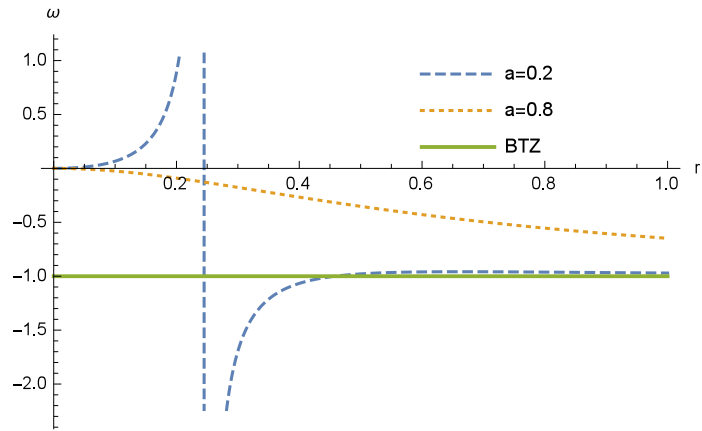}
    \caption{State parameter $\omega(r)$. For this plot we have considered $M=1$ and $l=0.5$.}
    \label{fig9}
\end{figure}

Finally, it is important to investigate the state function $\omega$ by considering a linear equation of state $p_r=\omega(r)\rho$ in order to study what kind of matter would sustain such black holes or wormholes. The state function is given by
\begin{eqnarray}\label{omega}
    \omega(r)=\frac{p_r}{\rho}=-\frac{r^2 \left(a^2+r^2\right)}{a^4+a^2 \left(2 r^2-l^2 M\right)+r^4}.
\end{eqnarray}
From the above equation we can notice that in the limits when $a\rightarrow0$ or $r\rightarrow\infty$ the state function goes to $\omega=-1$ (Dark Energy), which is the usual result for the BTZ black hole. Moreover, the state function can only be positive if the denominator is negative and this occurs only if $0<r<\sqrt{a l \sqrt{M}-a^2}$. In fig.(\ref{fig9}) we depict the state function behaviour in terms of the radius. We have considered $M=1$ and $l=0.5$. For this choice of parameters $a=0.2$ describes a black hole, and we can see that there is a region where it is possible to source this black hole with ordinary matter ($0<\omega<1$). Besides, after the discontinuity in the state function given by $\rho=0$, we have regions sourced by phantom ($\omega<-1$). For $a=0.8$ we are in the traversable wormhole domain. For this case we can clearly see from (\ref{fig9}) that the state function is always negative, hence the Simpson-Visser (2+1) dimensional wormhole can never be sourced by ordinary matter. 

\section{Thermodynamics}

In possession of the solution for the Simpson-Visser BTZ black hole given by (\ref{metric}), we are able to study the thermodynamics of the black hole by computing the Hawking’s temperature by means of $T_H=\frac{f'(r_h)}{4\pi}$, where $r_h$ is the radius of the horizon. The Hawking temperature for the present model exhibit no modifications in comparison to the usual BTZ black hole case, i.e., the Hawking temperature is given by $T_H=\frac{r_h}{2 \pi  l^2}$, showing the usual linear behaviour with the horizon radius. 

The entropy of the Simpson-Visser BTZ black hole remains also unmodified in comparison to the usual BTZ black hole. Thus the entropy, given by $dS=dM/T_H$ yields for the present context the usual relation $S=4\pi r_h$, i.e., twice the perimeter of the black hole. The same occurs with the specific heat, given by $C_v=dM/dT_H$, which yields the usual result $C_v=4\pi r_h$. Therefore we can conclude that the proposed modification in the BTZ black hole leads to practically no modification in the thermodynamic parameters. 

\begin{figure}[ht!]
    \centering
    \includegraphics[scale=0.7]{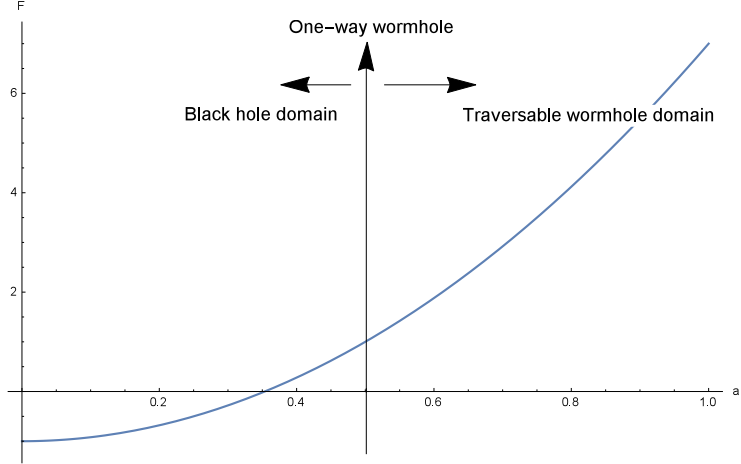}
    \caption{Free energy as a function of the regulator parameter $a$. For this plot we have considered $l=0.5$ and $M=1$.}
    \label{fig2}
\end{figure}

However the free energy of the system is slightly changed. Being the free energy given by
\begin{eqnarray}
F=M-T_H S,
\end{eqnarray}
we obtain for our case 
\begin{eqnarray}
F=\frac{(a^2-r_h^2)}{l^2}.
\end{eqnarray}
It is convenient to write the above expression for the free energy in terms of the mass $M$ and the cosmological constant, in order to properly analyze the domains of black hole and wormhole discussed previously. So that,
\begin{eqnarray}\label{free}
    F=\frac{2 a^2-l^2 M}{l^2}.
\end{eqnarray}
From eq. (\ref{free}) it is clear that $F>0$ ($F<0$) if $a^2>l^2M/2$ ($a^2<l^2M/2$), and $F=0$ when $a^2=l^2M/2$. We can see the behaviour of the free energy as a function of the regulator parameter $a$ depicted in fig(\ref{fig2}). We have considered $l=0.5$ and $M=1$ for the plot of fig. (\ref{fig2}). We can see that the black hole domain admits both positive and negative values for the free energy, depending on the value of the regulator parameter $a$. On the other hand, the one-way wormhole and the traversable wormhole domain exhibit only positive values for the free energy.

\section{orbits analysis}

Now, let us study the behaviour of a test particle around the present modified BTZ black hole. The particle's geodesic in orbit around a static black hole is given by
\begin{eqnarray}
    \dot{r}^2=\omega^2-f(r)\left(\frac{L^2}{r^2+a^2}+\epsilon\right),
\end{eqnarray}
where $\omega$ is the particle's energy, $L$ is the angular momentum and $\epsilon=\{0,+1\}$ defines massive ($\epsilon=+1$) particles or massless ($\epsilon=0$) particles. Thus the effective potential is written as
\begin{eqnarray}\label{effectivepot}
    V_{\epsilon}(r)=f(r)\left(\frac{L^2}{r^2+a^2}+\epsilon\right).
\end{eqnarray}
The circular geodesics occur at the points $r$ satisfying $V_{\epsilon}'(r)=0$. Now we will analyze the orbits for both massless and massive particles. 

For massless particles we have the circular orbits occurring at $V_0'(r)=0$, i.e., 
\begin{eqnarray}
    V_0'(r)=\frac{2 L^2 M r}{\left(a^2+r^2\right)^2}.
\end{eqnarray}
Therefore, for massless particles the only solution for $V_0'(r)=0$ occurs at $r=0$. This solution must be disregarded as a physical solution in the present context since this spherical surface is not valid for the location of a photon sphere. 

Analogously, for the case of massive particles we have the circular orbits occurring at $V_0'(r)=0$. From eq.(\ref{effectivepot}) we obtain 
\begin{eqnarray}
    V_{1}'(r)=2 r \left(\frac{L^2 M}{\left(a^2+r^2\right)^2}+\frac{1}{l^2}\right).
\end{eqnarray}
Clearly, we have one solution for $V_0'(r)=0$ at $r=0$, which is not valid as an orbit solution for the same reason discussed previously. Moreover, we have four non-trivial solutions, but all of them are complex valued solutions. This means that these solutions are not physical, since the $r$ must be a real valued quantity. 

\begin{figure*}[h!]
    \centering
\begin{subfigure}{.5\textwidth}
  \centering
  \includegraphics[scale=0.8]{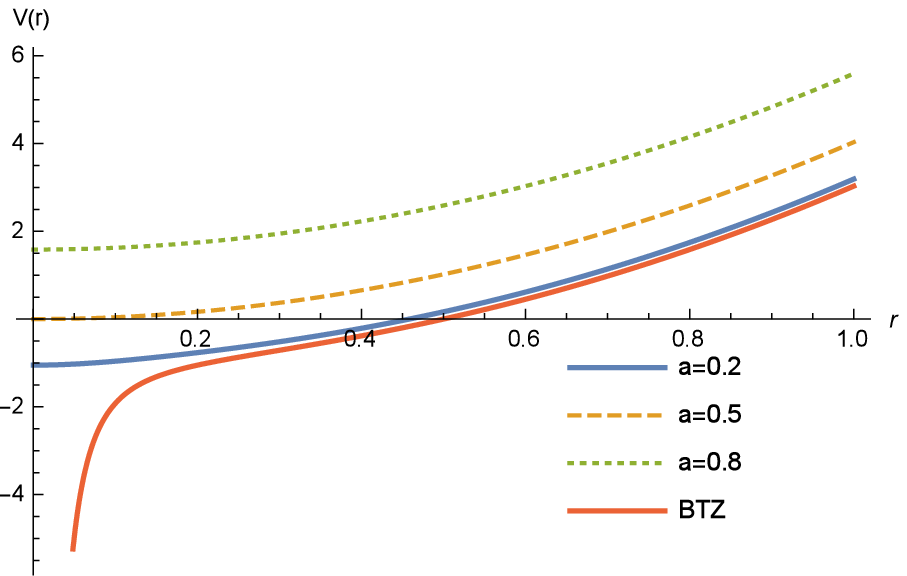}
  \caption{}
 
\end{subfigure}%
\begin{subfigure}{.5\textwidth}
  \centering
  \includegraphics[scale=0.8]{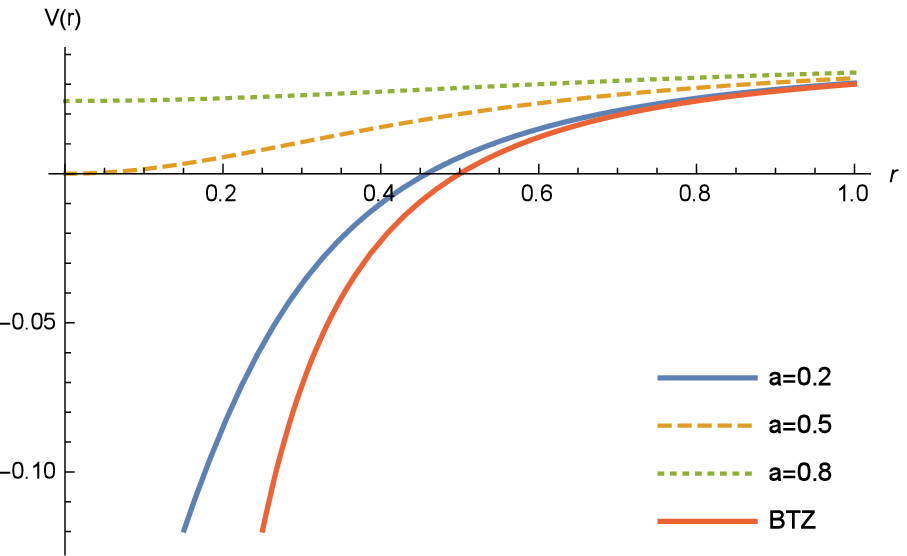}
  \caption{}

\end{subfigure}%
    \caption{Effective potential (a) for massive particles and (b) for massless particles. For this plot we have considered, $l=0.5$, $M=1$ and $L=0.1$.}
    \label{fig3}
\end{figure*}

In Fig. (\ref{fig3}) we depict the effective potential of massless and massive particles for the Simpson-Visser BTZ black hole. We have considered for this plot $l=0.5$, $M=1$ and $L=0.1$, so that for this case $a<0.5$ is a black hole, $a=0.5$ is a one-way wormhole and $a>0.5$ is a traversable wormhole. 

\section{Simpson-Visser Charged BTZ black hole}

\begin{figure*}[h!]
    \centering
\begin{subfigure}{.5\textwidth}
  \centering
  \includegraphics[scale=0.8]{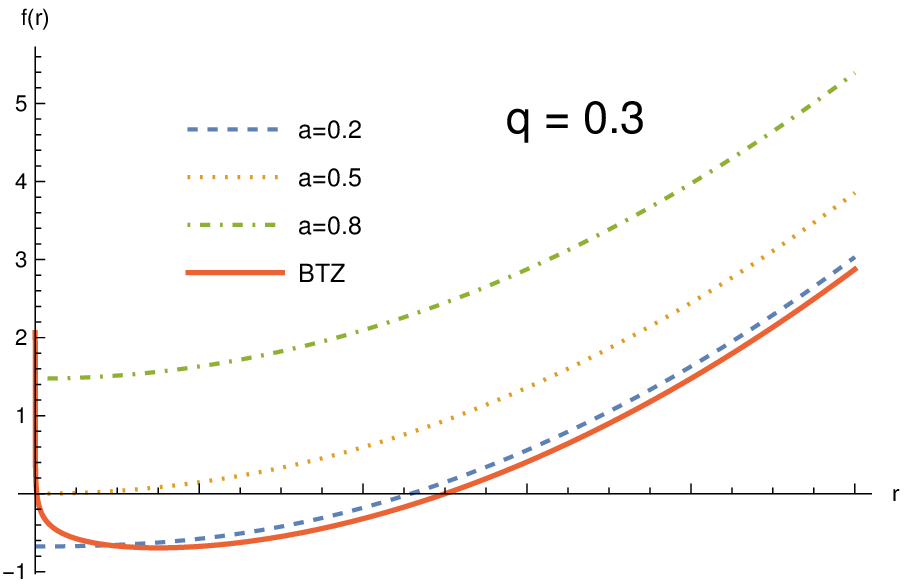}
  \caption{}

\end{subfigure}%
\begin{subfigure}{.5\textwidth}
  \centering
  \includegraphics[scale=0.8]{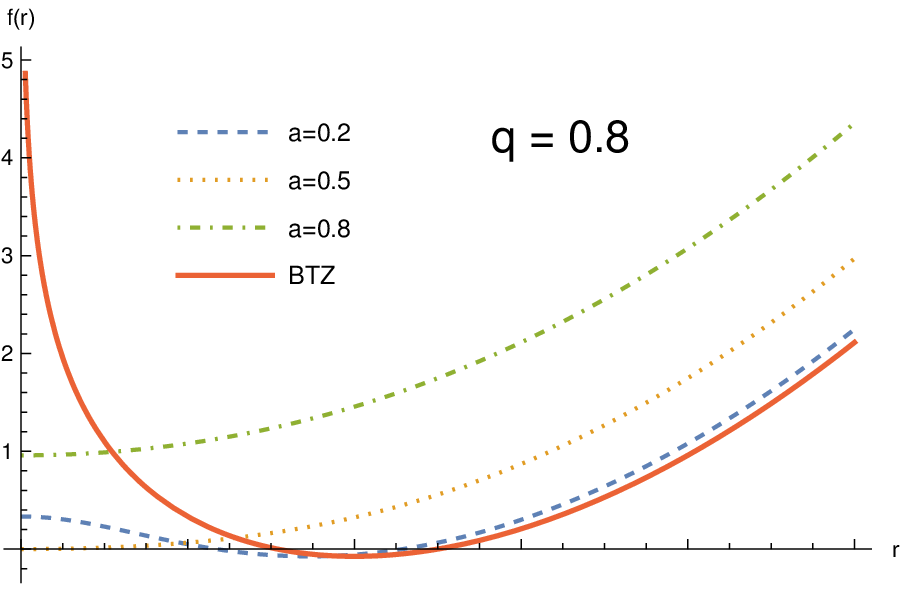}
  \caption{}

\end{subfigure}%

    \caption{Plot of $f(r)$ for the Simpson-Visser charged BTZ black hole. We have considered for this plot $M=1$ and $l=0.5$.}
    \label{fig4}
\end{figure*}

Soon after the Simpson-Visser black-bounce was proposed, the charged version was studied in Ref. \cite{Franzin:2021vnj}. The authors found a solution that interpolates between a regular charged black hole and a charged wormhole.  Here, we will analyze the case of a charged BTZ black hole under the Simpson-Visser regularization. For this case the metric is given by (\ref{metric}) but with

\begin{eqnarray}\label{f_r_charged}
    f(r)=-M+\frac{a^2+r^2}{l^2}-2 q^2 \ln \left(\frac{\sqrt{a^2+r^2} }{l}\right),
\end{eqnarray}
 Obviously, when $a=0$, we recover the usual expression of $f(r)$ for the charged BTZ black hole. The location of the event horizon $r_h$ can be obtained by $f(r_h)=0$, however analytic solutions for $r_h$ are not easily reachable since $f(r)=0$ yields a transcendental equation for $r_h$. In order to have a clearer view of the Simpson-Visser charged BTZ black hole, we have plotted the behaviour of the $f(r)$ in fig.(\ref{fig4}) for two configurations of charge, namely, $q=0.3$ and $q=0.8$.  

From the horizon curves we can obtain the mass of the black hole in terms of the horizon radius as
\begin{eqnarray}
    M=\frac{a^2+r_h^2}{l^2}-2 q^2 \log \left(\frac{\sqrt{a^2+r_h^2}}{l}\right).
\end{eqnarray}
A similar analysis gives us the Hawking temperature for the present case. Differently from the uncharged case, the Simpson-Visser charged BTZ black hole exhibit a different Hawking temperature, given by
\begin{eqnarray}
    T_H=\frac{r_h}{2 \pi  l^2}-\frac{q^2 r_h}{2 \pi  (a^2+r_h^2)}.
\end{eqnarray}
The Hawking temperature for the usual charged BTZ black hole is in fact divergent at $r=0$. The Simpson-Visser regularization provides a modification in the charge dependent term which indeed removes the divergence. Moreover the Hawking temperature vanishes for $r_h=0$ and $r_h=\pm\sqrt{l^2 q^2-a^2}$. We must select the positive solutions, since they provide the possibility of a non-negative horizon radius. 

\begin{figure*}[h!]
    \centering
\begin{subfigure}{.5\textwidth}
  \centering
  \includegraphics[scale=0.8]{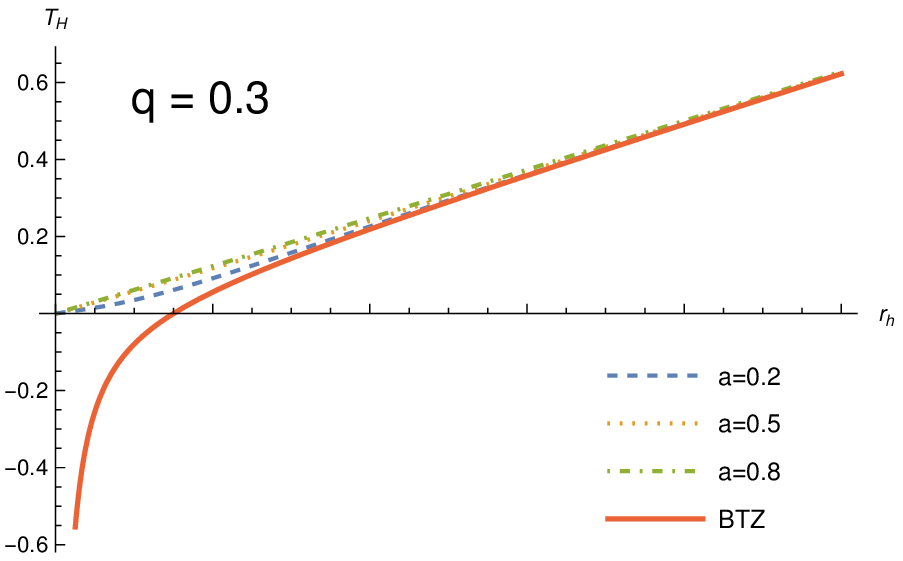}
  \caption{}

\end{subfigure}%
\begin{subfigure}{.5\textwidth}
  \centering
  \includegraphics[scale=0.8]{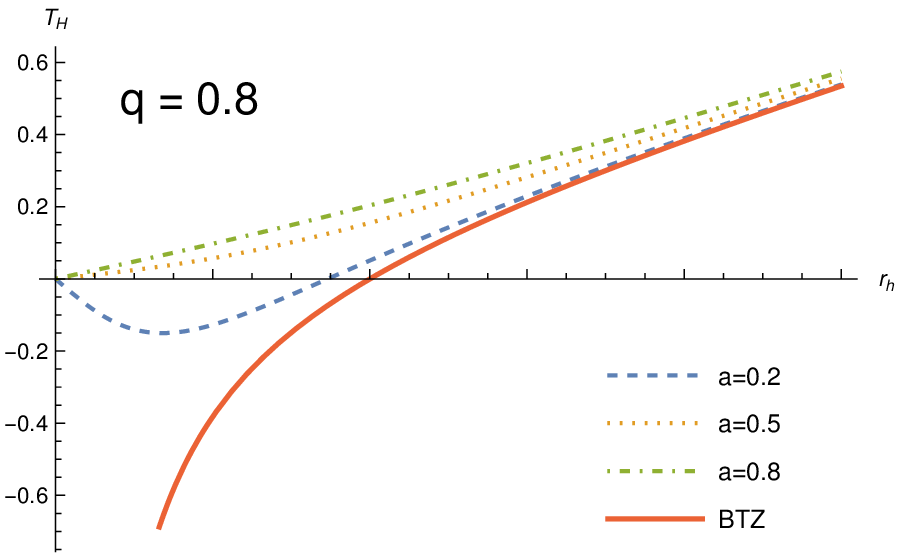}
  \caption{}

\end{subfigure}%

    \caption{Plot of the Hawking temperature for the Simpson-Visser charged BTZ black hole. We have considered for this plot $M=1$ and $l=0.5$.}
    \label{fig5}
\end{figure*}

In fig.(\ref{fig5}) we depicted the behaviour of the Hawking temperature for the charged Simpson-Visser BTZ black hole for $q=0.3$ and $q=0.8$. As we can see, for any $a\neq0$ case we have a vanishing Hawking temperature at the origin and for large $r_h$ the Hawking temperature recovers the linear behaviour $T_H=\frac{r_h}{2 \pi  l^2}$. In other words, for large $r_h$ the Hawking temperature no longer sees the presence of the charge. 

\begin{figure*}[h!]
    \centering
\begin{subfigure}{.5\textwidth}
  \centering
  \includegraphics[scale=0.8]{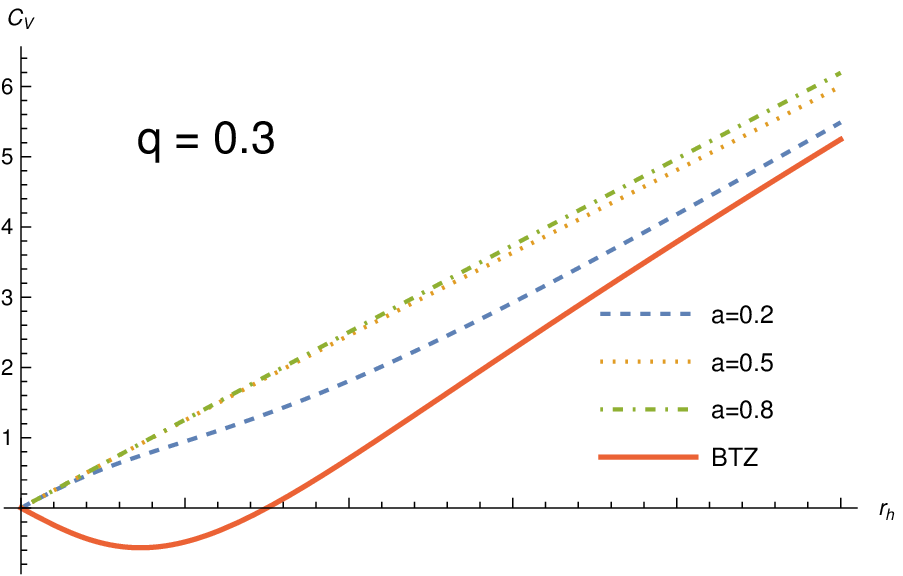}
  \caption{}

\end{subfigure}%
\begin{subfigure}{.5\textwidth}
  \centering
  \includegraphics[scale=0.8]{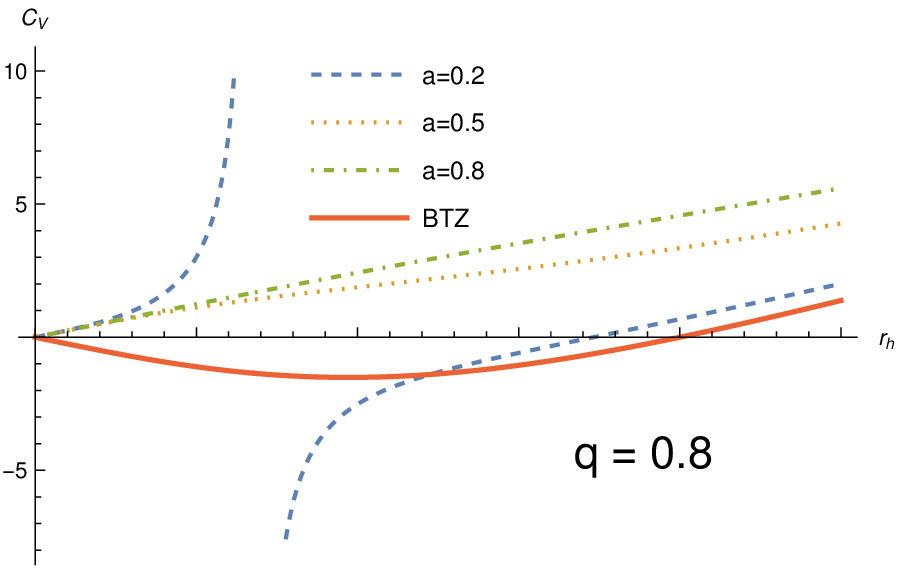}
  \caption{}

\end{subfigure}%

    \caption{Plot of the heat capacity for the Simpson-Visser charged BTZ black hole. We have considered for this plot $l=0.5$.}
    \label{fig6}
\end{figure*}

The entropy for the Simpson-Visser charged BTZ black hole, similarly to the uncharged case, presents no modification due to the regularization. Hence it yields the usual result $S=4\pi r_h$. However the specific heat is modified by the Simpson-Visser regularization, so that, 
\begin{eqnarray}\label{chargedheat}
    C_v=4\pi r_h-\frac{8\pi l^2 q^2 r_h^3}{a^4+a^2 \left(2 r_h^2-l^2 q^2\right)+l^2 q^2 r_h^2+r_h^4}.
\end{eqnarray}
As it is widely known, the thermodynamical stability of black holes (BTZ black holes for our case) is directly related to the sign of the heat capacity. A positive heat capacity indicates that the system is thermodynamically stable, while its negativity imply a thermodynamical instability. It is possible to obtain the points where eq.(\ref{chargedheat}) vanish. Such points are
\begin{eqnarray}
    r_h&=&0,\\
    r_h&=&\pm\sqrt{l^2 q^2-a^2},\\\
    r_h&=&\pm i a.
\end{eqnarray}
We must disregard the complex solutions and the negative ones, therefore we have only two physical solutions, namely, $r_h=0$ and $r_h=\sqrt{l^2 q^2-a^2}$. Moreover, notice that the location of $r_h$ is dependent on the bounce parameter $a$. Also, it is clear that we have a divergence in the specific heat when
\begin{eqnarray}\label{eq26}
    r_h=\pm\frac{\sqrt{\pm\sqrt{l^2 q^2 \left(8 a^2+l^2 q^2\right)}-2 a^2-l^2 q^2}}{\sqrt{2}}.
\end{eqnarray}
The divergence in the specific heat indicates a phase transition. As we can see, such phase transition appears as a relation between the bounce parameter and the electric charge. Since $r_h$ must be a real valued quantity, from eq.(\ref{eq26}) we reach at the condition $a<lq$ for the existence of a phase transition. The behaviour of the specific heat can be see in fig.(\ref{fig6}). We have considered two values for the charge, namely, $q=0.3$ and $q=0.8$, and $l=0.5$. 

The free energy is also modified by the bounce parameter, so that
\begin{eqnarray}
    F=\frac{a^4+2 l^2 q^2 r_h^2-r_h^4}{l^2 \left(a^2+r_h^2\right)}-2 q^2 \log \left(\frac{\sqrt{a^2+r_h^2}}{l}\right).
\end{eqnarray}
The divergence in the free energy when $r_h=0$ is clearly regularized by the bounce parameter. Besides, the usual charged BTZ result is recovered when $a=0$.

\begin{figure*}[h!]
    \centering
\begin{subfigure}{.5\textwidth}
  \centering
  \includegraphics[scale=0.8]{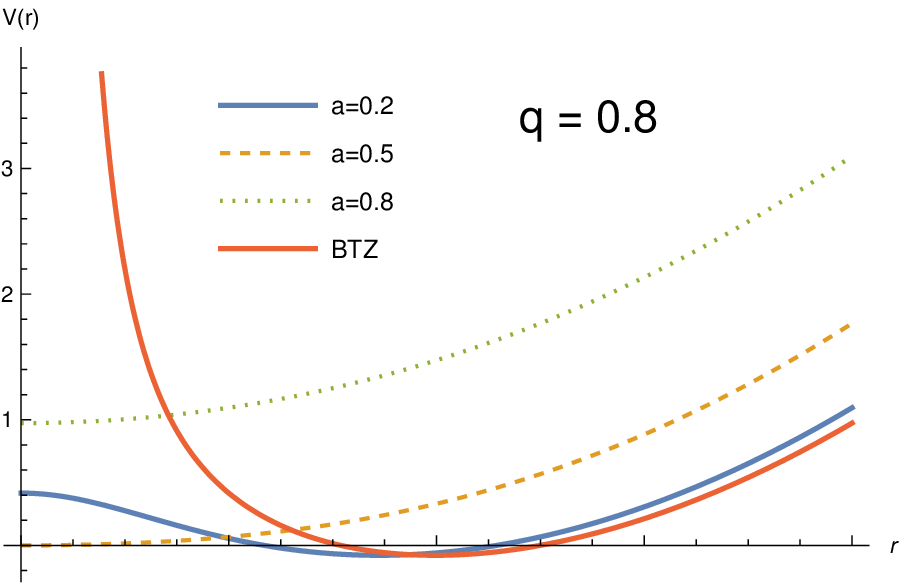}
  \caption{}

\end{subfigure}%
\begin{subfigure}{.5\textwidth}
  \centering
  \includegraphics[scale=0.8]{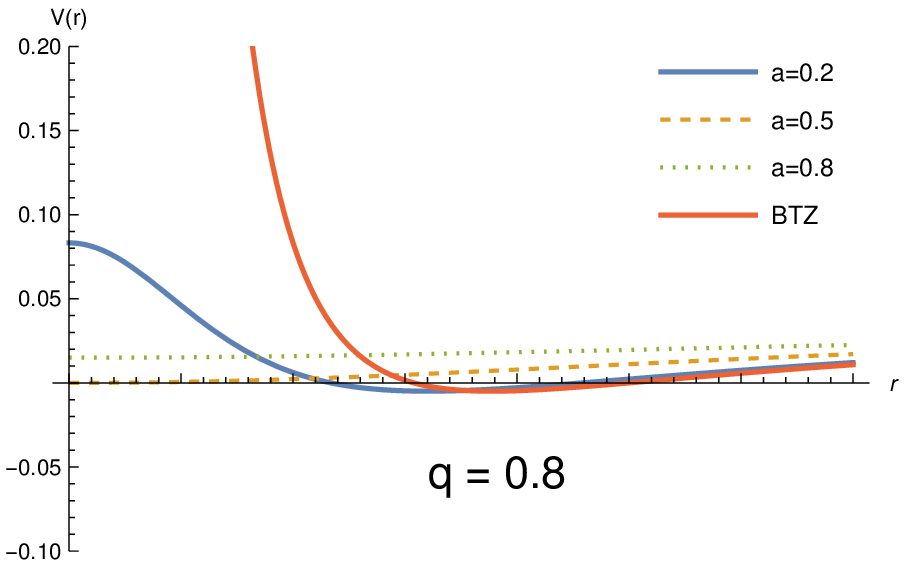}
  \caption{}

\end{subfigure}%

    \caption{Plot of the effective potential for the Simpson-Visser charged BTZ black hole. The left panel stands for massive particles while the right panel is for photons. We have considered for this plot $M=1$, $l=0.5$ and $L=0.1$.}
    \label{fig7}
\end{figure*}

Finally, the same analysis performed in the previous section for the orbits of massive and massless particles moving around the uncharged BTZ black hole can be done for the present case as well. By taking the effective potential described by eq(\ref{effectivepot}) and replacing $f(r)$ by eq.(\ref{f_r_charged}) we can discuss the orbits for the charged case. For both massive and massless cases exact solutions for the point where these orbits become stable couldn't be found. However graphically it is possible to see that there are stable orbits for both massive and massless cases. Moreover, the fig.(\ref{fig7}) allows us to see that the points where we have stable orbits are sensitive to the bounce parameter and charge. 

\section{Conclusion}\label{Sec-6}
In this manuscript, we apply the procedure proposed by Simpson and Visser to obtain a solution that interpolates between a regular BTZ  black hole and a wormhole.  Beyond being simple, it opens a new avenue to shed some light on the properties of both: wormholes and regular black holes. We have considered the charged and uncharged BTZ black holes. 

For the uncharged case, we first point out that the standard BTZ is already regular.  However, when we considered the parameter $a$, we found a regular BTZ black hole with only one horizon. This is very different from the Schwarzschild case, with two horizons with radius controlled by the parameter $a$. We also found that the entropy and Hawking temperature does not depends on $a$ and therefore the behavior is very similar to the standard BTZ. We also find that stable orbits are not allowed.  Finally, we compute the curvature tensors. We find that they are identical to the standard BTZ at $r=0$ and $r=\infty$. For other values of $r$ we get non-constant values. Therefore we get an effective source of the BTZ black hole that does not change its main properties, such as entropy. We have found, by studying the energy conditions and the state parameter, that there is a small range of parameters where we can source a black hole with ordinary matter, but wormholes are always sourced by exotic matter. 

Next, we study the charged case.  In this case, the BTZ black hole is not regular, and the introduction of the parameter $a$ regularizes it. Differently from the uncharged case, the charged case exhibits modifications due to the parameter $a$. We could verify that for the charged case, the bounce parameter $a$ affects the point where the Hawking temperature vanishes. The heat capacity for the charged case is drastically modified by the Simpson-Visser regularization. We could verify that the when $a<lq$ we have a phase transition where the heat capacity changes sign, affecting therefore the stability of the black hole. Moreover, graphically we could see that there are stable orbits for both massive and massless cases. The fig.(\ref{fig7}) allows us to see that the points where we have stable orbits are sensitive to the bounce parameter and charge.

Finally, we should point out that, as far as we know, this is the first time that the Simpson and Visser procedure is applied to dimensions fewer than $3+1$. This opens new avenues of study, including cases with dimensions higher than $3+1$. This is the subject of future study by us. 

\section*{Acknowledgement}
The authors would like to thanks Alexandra Elbakyan and sci-hub, for removing all barriers
in the way of science. We acknowledge the financial support provided by the Coordenação de Aperfeiçoamento de Pessoal de Nível Superior or (CAPES), the Conselho Nacional de Desenvolvimento Científico e Tecnológico (CNPq) and Fundaçao Cearense de Apoio ao Desenvolvimento Científico e
Tecnológico (FUNCAP) through PRONEM PNE0112- 00085.01.00/16.

\end{document}